\documentclass{LMCS}

\usepackage[latin1]{inputenc}
\usepackage{amssymb}
\usepackage{enumerate}
\usepackage{hyperref}

\newtheorem{Rem}[thm]{Remark}
\theoremstyle{plain}
\theoremstyle{plain}\newtheorem{Pro}[thm]{Proposition}
\theoremstyle{plain}

\def\ufootnote#1{\let\savedthfn\thefootnote\let\thefootnote\relax
\footnote{#1}\let\thefootnote\savedthfn\addtocounter{footnote}{-1}}

\newcommand{\bormone}{{\bf\Pi}^{0}_{1}}

\newcommand{\bormtwo}{{\bf\Pi}^{0}_{2}}

\newcommand{\bormom}{{\bf\Pi}^{0}_{\omega}}
\newcommand{\borom}{{\bf\Delta}^{0}_{\omega}}

\newcommand{\borapxi}{{\bf\Sigma}^{0}_{\xi}}

\newcommand{\bormpxi}{{\bf\Pi}^{0}_{\xi}}

\newcommand{\borpxi}{{\bf\Delta}^{0}_{\xi}}
\newcommand{\borel}{{\bf\Delta}^{1}_{1}}

\newcommand{\borone}{{\bf\Delta}^{0}_{1}}
\newcommand{\bortwo}{{\bf\Delta}^{0}_{2}}

\newcommand{\boraone}{{\bf\Sigma}^{0}_{1}}
\newcommand{\boratwo}{{\bf\Sigma}^{0}_{2}}

\newcommand{\boraom}{{\bf\Sigma}^{0}_{\omega}}

\newcommand{\hs}{\relax}

\newcommand{\noi}{\relax}

\newcommand{\om}{\omega}
\newcommand{\Si}{\Sigma}
\newcommand{\Sis}{\Sigma^\star}
\newcommand{\Sio}{\Sigma^\omega}

\newcommand{\lra}{\leftrightarrow}
\newcommand{\fa}{\forall}
\newcommand{\ra}{\rightarrow}

\newcommand{\Ga}{\Gamma}

\newcommand{\Gao}{\Gamma^\omega}
\newcommand{\ite}{\item}

\newcommand{\ol}{$\omega$-language}

\def\doi{5 (4:4) 2009}
\lmcsheading%
{\doi}
{1--18}
{}
{}
{Jan.~20, 2009}
{Dec.~21, 2009}
{}   

\begin{document}

\title{The  Complexity of Infinite Computations  In   Models of Set Theory} 

\author{Olivier Finkel} 
\address{Equipe de Logique Math\'ematique
 \\  CNRS et Universit\'e Paris 7, France.}
\email{finkel@logique.jussieu.fr}


\begin{abstract}
We prove the following surprising result: there exist a $1$-counter
B\"uchi automaton and a $2$-tape B\"uchi automaton such that the
$\om$-language of the first and the infinitary rational relation of
the second in one model of {\bf ZFC} are ${\bf \Pi}_2^0$-sets, while
in a different model of {\bf ZFC} both are analytic but non Borel
sets.

This shows that the topological complexity of an $\om$-language
accepted by a $1$-counter B\"uchi automaton or of an infinitary
rational relation accepted by a $2$-tape B\"uchi automaton is not
determined by the axiomatic system {\bf ZFC}.

We show that a similar result holds for the class of languages of
infinite pictures which are recognized by B\"uchi tiling systems.

We infer from the proof of the above results an improvement of the
lower bound of some decision problems recently studied by the author.
\end{abstract}

\maketitle

\section{Introduction}

\noindent  Acceptance of infinite words by finite automata was firstly 
considered in the sixties by B\"uchi in order to study 
the decidability of the monadic second order theory 
of one successor over the integers \cite{Buchi62}. The class of regular $\om$-languages  has been 
intensively studied and many applications have been found, see \cite{Thomas90,Staiger97,PerrinPin} for many results and references. 
 Many extensions of regular $\om$-languages have been investigated as the classes of  $\om$-languages
accepted by $1$-counter automata, pushdown automata, $2$-tape automata, Petri nets, Turing machines, 
see \cite{Thomas90,eh,Staiger97,Fin-survey} for a survey of this work. 

\hs  A way to study the complexity of   languages of infinite words  
accepted by finite machines          is to study    their      topological complexity       and firstly 
to locate them with regard to 
the Borel and the projective hierarchies. 
This work was analysed in \cite{Staiger86a,Staiger87b,Thomas90,Simonnet92,eh,LescowThomas,Staiger97}.  
It is well known that every $\om$-language accepted by a deterministic B\"uchi automaton is a ${\bf \Pi}^0_2$-set. This implies  that 
  any $\om$-language accepted by a deterministic   Muller automaton is a 
boolean combination of ${\bf \Pi}^0_2$-sets hence a
${\bf \Delta}^0_3$-set.       But then it follows from Mc Naughton's Theorem, that 
all regular $\om$-languages, which are accepted by deterministic Muller automata, are also 
${\bf \Delta}^0_3$-sets. 
The Borel hierarchy of regular $\om$-languages is then determined. 
Moreover Landweber proved that one can effectively determine  
the Borel complexity of a  regular $\om$-language accepted by a given  Muller or B\"uchi  automaton,  see \cite{Landweber69,Thomas90,Staiger97,PerrinPin}. 

\hs In recent papers \cite{Fin-mscs06,Fink-Wd}  we have proved the following very surprising results. 
From the topological point of view,  $1$-counter B\"uchi automata  and  $2$-tape B\"uchi automata  have the same accepting power as 
Turing machines equipped with a  B\"uchi acceptance condition. 
In particular,   for every non null recursive ordinal $\alpha$,  
there exist some  
${\bf \Si}^0_\alpha$-complete and some ${\bf \Pi}^0_\alpha$-complete $1$-counter  $\om$-languages (respectively,   infinitary rational relations).  
 And the supremum of the set of Borel ranks of $1$-counter  $\om$-languages (respectively,  
infinitary rational relations) is an ordinal $\gamma_2^1$ which is strictly greater than the first non recursive ordinal   
$\om_1^{\mathrm{CK}}$.   Moreover we have proved that there is no general algorithm to determine in an effective way the topological complexity 
of a given $1$-counter  $\om$-language (respectively,  infinitary rational relation).  Topological properties of $1$-counter  $\om$-languages 
(respectively,   infinitary rational relations) are actually highly undecidable: for any countable ordinal $\alpha$, 
``determine whether a given $1$-counter  $\om$-language (respectively,  infinitary rational relation) is in the Borel class ${\bf \Si}^0_\alpha $ (respectively, 
${\bf \Pi}^0_\alpha $)" is a $\Pi^1_2$-hard problem, \cite{Fin-HI}. 
 
\hs We prove here an even more amazing result which shows that Set Theory is actually very important  in the study of infinite computations. 
Recall that the usual axiomatic system {\bf ZFC} is 
Zermelo-Fraenkel system {\bf ZF}   plus the axiom of choice {\bf AC}.
We prove that there  exist a $1$-counter 
B\"uchi automaton $\mathcal{A}$ and a $2$-tape B\"uchi automaton  $\mathcal{B}$ such that :
\begin{enumerate}[(1)]
\item There is a  model $V_1$ of {\bf ZFC} in which the  $\om$-language 
$L(\mathcal{A})$ and the infinitary rational relation $L(\mathcal{B})$ are   ${\bf \Pi}_2^0$-sets, and 
\item  There is a  model  $V_2$ of {\bf ZFC}  in which 
the $\om$-language  $L(\mathcal{A})$ and the infinitary rational relation $L(\mathcal{B})$ are analytic but non Borel sets. 
\end{enumerate}
 This shows that  the topological complexity of an $\om$-language accepted by a $1$-counter B\"uchi automaton or of an infinitary rational relation 
accepted by a $2$-tape B\"uchi automaton  
is not determined by the  axiomatic system {\bf ZFC}.

 We show that a similar result holds for the class of languages of
 infinite pictures which are recognized by B\"uchi tiling systems,
 recently studied by Altenbernd, Thomas and W\"ohrle in \cite{ATW02},
 see also \cite{Finkel04,Fink-tilings}.

\hs In order to prove these results, we consider the largest thin (i.e., without perfect subset) effective coanalytic subset of the Cantor space $2^\om$. 
The existence of this largest thin $\Pi_1^1$-set $\mathcal{C}_1$ was proven by Kechris in \cite{Kechris75} and independently 
by Guaspari  and  Sacks in \cite{Guaspari,Sacks}. By considering the cardinal of this set $\mathcal{C}_1$ in different models of set theory, we 
show  that  its topological complexity depends on the actual model of {\bf ZFC}. Then we use some constructions from recent papers 
\cite{Fin-mscs06,Fin06b,Fink-tilings} to infer our new results about $1$-counter or 
$2$-tape  B\"uchi   automata and B\"uchi tiling systems. 
From the proof of the above results and from Shoenfield's Absoluteness Theorem we get 
an improvement of  the lower bound of some decision problems recently studied in \cite{Fin-HI,Fink-tilings}. 
We show that the problem to determine whether an  $\om$-language accepted by a given real time $1$-counter B\"uchi automaton 
(respectively, an  infinitary rational relation 
accepted by a given  $2$-tape B\"uchi automaton) is in the 
Borel class ${\bf \Si}^0_\alpha $ (respectively, 
${\bf \Pi}^0_\alpha $), for a countable ordinal $\alpha>2$ (respectively, $\alpha\geq 2$),  
is not  in the class $ \Pi_2^1$. A similar result holds for 
 languages of infinite pictures accepted by  B\"uchi tiling systems.

\hs The paper is organized as follows. In Section 2 we recall definitions of counter automata, $2$-tape automata, and tiling systems. 
 We recall basic notions of topology in Section 3.   Results on the  largest effective coanalytic set are stated in Section 4. 
We prove our main results in Section 5. 

\hs Notice that as  the results presented in this paper might be of interest to  both set theorists and theoretical computer scientists, we 
shall recall in detail in Section 2 some notions of automata theory which are well known to computer scientists but not to set theorists. In a similar 
way we give in Sections 3 and 4 a presentation of  some  results of set theory which are well known to  set theorists   but not to      computer scientists.

\section{Automata}
 
\noindent We assume now  the reader to be familiar with the theory of formal $\om$-languages  
\cite{Thomas90,Staiger97}.
We shall follow usual notations of formal language theory. 

 When $\Si$ is a finite alphabet, a {\it non-empty finite word} over $\Si$ is any 
sequence $x=a_1\ldots a_k$, where $a_i\in\Sigma$ 
for $i=1,\ldots ,k$ , and  $k$ is an integer $\geq 1$. The {\it length}
 of $x$ is $k$, denoted by $|x|$.
 The {\it empty word} has no letter and is denoted by $\lambda$; its length is $0$. 
 $\Sis$  is the {\it set of finite words} (including the empty word) over $\Sigma$.

  The {\it first infinite ordinal} is $\om$.
 An $\om$-{\it word} over $\Si$ is an $\om$ -sequence $a_1 \ldots a_n \ldots$, where for all 
integers $ i\geq 1$, ~
$a_i \in\Sigma$.  When $\sigma$ is an $\om$-word over $\Si$, we write
 $\sigma =\sigma(1)\sigma(2)\ldots \sigma(n) \ldots $,  where for all $i$,~ $\sigma(i)\in \Si$,
and $\sigma[n]=\sigma(1)\sigma(2)\ldots \sigma(n)$  for all $n\geq 1$ and $\sigma[0]=\lambda$.

 The usual concatenation product of two finite words $u$ and $v$ is 
denoted $u.v$ (and sometimes just $uv$). This product is extended to the product of a 
finite word $u$ and an $\om$-word $v$: the infinite word $u.v$ is then the $\om$-word such that:
\[(u.v)(k)=u(k)\quad\hbox{if $k\leq |u|$, and}\quad
  (u.v)(k)=v(k-|u|)\quad\hbox{if $k>|u|$.}
\]
 The {\it set of } $\om$-{\it words} over  the alphabet $\Si$ is denoted by $\Si^\om$.
An  $\om$-{\it language} over an alphabet $\Sigma$ is a subset of  $\Si^\om$.  The complement (in $\Sio$) of an 
$\om$-language $V \subseteq \Sio$ is $\Sio - V$, denoted $V^-$.

 For a finitary language $V \subseteq \Si^{\star}$, the $\om$-power of $V$ is the $\om$-language 
 $$V^\om = \{ u_1\ldots  u_n\ldots  \in \Si^\om \mid  \fa i\geq 1 ~~ u_i\in V \}$$

\hs Abstract models of finite machines reading  finite or infinite words have been considered in automata theory, calculability and complexity 
theories. The simplest model of machine used for recognizability of  languages of  (finite or infinite) words is the model of finite state machine. 
One can consider that such a machine $\mathcal{M}$ has a semi infinite tape divided into cells. This tape contains at the beginning the input word written 
from left to right,  each letter being contained in a cell; in the case of a finite input word, the remaining cells contain a special blank symbol. The machine has a 
reading (only) head, placed at the beginning on the first cell. It has also a finite control, consisting of a finite set $K$ of states and a current state. There is a special 
state $q_0$ called the initial state and a set  $F \subseteq K$ of final states. 
The reading of a word  begins in state $q_0$; then the machine reads successively the letters from left to right, 
changing the  current state according to the transition relation  which has a finite description. The finite word $x$ is accepted by $\mathcal{M}$ if the reading 
of $x$ ends in a final state. An infinite word $\sigma$ is accepted by $\mathcal{M}$ if some final state occurs infinitely often during the reading 
of $\sigma$. We now give a formal definition of a finite state machine.

\begin{defi} 
A finite state machine (FSM) is a quadruple $\mathcal{M}=(K,\Si,\delta, q_0)$, where $K$ 
is a finite set of states, $\Sigma$ is a finite input alphabet, $q_0 \in K$ is the initial state
and $\delta$ is a mapping from $K \times   \Si$ into $2^K$. 

 Let $x =a_1a_2\ldots a_n$ be a  finite word over $\Si$.  
 A sequence of states $r=q_1q_2\ldots q_n q_{n+1}$  is called a run of $\mathcal{M}$ on $x$ 
 iff:
\begin{enumerate}[(1)]
\item $q_1=q_0$ is the initial state, and 
\item for each $i\geq 1$, $q_{i+1} \in \delta( q_i,a_i)$.
\end{enumerate}
Let $\sigma =a_1a_2\ldots a_n\ldots$ be an  $\om$-word over $\Si$.
 A sequence of states $r=q_1q_2\ldots q_n\ldots$  is called an (infinite) run of $\mathcal{M}$ on $\sigma$ 
 iff:
\begin{enumerate}[(1)]
\item $q_1=q_0$ is the initial state, and 
\item for each $i\geq 1$, $q_{i+1} \in \delta( q_i,a_i)$.
\end{enumerate}
For every (infinite) run $r=q_1q_2\ldots q_n\ldots $ of $\mathcal{M}$, $In(r)$ is the set of
states  entered infinitely often by $\mathcal{M}$ during the run $r$. 
\end{defi}

\begin{defi}
An automaton  is a 5-tuple $\mathcal{M}=(K,\Si,\delta, q_0, F)$ where
  $\mathcal{M}'=(K,\Si,\delta, q_0)$
is a finite state machine and $F\subseteq K$ is the set of final states.
The language  accepted by $\mathcal{M}$ is the set of finite words $x$ such that there is a run of 
$\mathcal{M}$ on $x$ ending in a final state. 
\end{defi}

\begin{defi}
A B\"uchi automaton  is a 5-tuple $\mathcal{M}=(K,\Si,\delta, q_0, F)$ where
  $\mathcal{M}'=(K,\Si,\delta, q_0)$
is a finite state machine and $F\subseteq K$ is the set of final states.
The $\om$-language  accepted by $\mathcal{M}$ is 
$$L(\mathcal{M})= \{  \sigma\in\Si^\om  \mid  \mbox{ there exists a  run } r
\mbox{ of } \mathcal{M} \mbox{ on } \sigma \mbox{ such that } In(r) \cap F \neq\emptyset \}.$$
\end{defi}

\noi  Recall that a language (respectively, $\om$-language) is said to be regular iff it is accepted by an automaton 
(respectively, B\"uchi automaton).  An $\om$-language $L$ is regular iff it belongs to the $\om$-Kleene closure of the class of 
finitary regular languages, i.e. iff there exist some regular languages $U_i, V_i$, for $i\in [1, n] $,  such that 
$L = \bigcup_{i=1}^n U_i.V_i^\om$. 

\hs Notice that a finite state machine  has only a bounded memory containing the current state of the machine. 
More complicated machines have been considered which can store some unbounded contents. In particular a counter machine has a 
finite set of counters, each of which containing a  non-negative integer. The machine can test whether the content of a given counter is zero or not. 
And transitions depend on the letter read by the machine, the current state of the finite control, and the tests about the values of the counters. Each transition 
leads to another state, and values of the counters can be increased by  $1$ or decreased by $1$, providing that these values always remain  non-negatives.  
Notice that in this model some  $\lambda$-transitions are allowed. During these transitions the reading head of the machine does not move to the right, i.e. 
 the machine does not  read any more letter.

\hs  We now recall the formal definition of $k$-counter machine and $k$-counter B\"uchi automata which will be useful in the sequel. 

\begin{defi} Let $k$ be an integer $\geq 1$. 
A  $k$-counter machine  is a 4-tuple 
$\mathcal{M}$=$(K,\Si,$ $ \Delta, q_0)$,  where $K$ 
is a finite set of states, $\Sigma$ is a finite input alphabet, 
 $q_0\in K$ is the initial state, 
and  $\Delta \subseteq K \times ( \Si \cup \{\lambda\} ) \times \{0, 1\}^k \times K \times \{0, 1, -1\}^k$ is the transition relation. 
The $k$-counter machine $\mathcal{M}$ is said to be {\it real time} iff: 
$\Delta \subseteq K \times
  \Si \times \{0, 1\}^k \times K \times \{0, 1, -1\}^k$, 
 i.e. iff there are no  $\lambda$-transitions. 
  
If  the machine $\mathcal{M}$ is in state $q$ and 
$c_i \in \mathbb{N}$ is the content of the $i^{th}$ counter 
 $\mathcal{C}$$_i$ then 
the  configuration (or global state)
 of $\mathcal{M}$ is the  $(k+1)$-tuple $(q, c_1, \ldots , c_k)$.

\hs For $a\in \Si \cup \{\lambda\}$, 
$q, q' \in K$ and $(c_1, \ldots , c_k) \in \mathbb{N}^k$ such 
that $c_j=0$ for $j\in E \subseteq  \{1, \ldots , k\}$ and $c_j >0$ for 
$j\notin E$, if 
$(q, a, i_1, \ldots , i_k, q', j_1, \ldots , j_k) \in \Delta$ where $i_j=0$ for $j\in E$ 
and $i_j=1$ for $j\notin E$, then we write:
$$a: (q, c_1, \ldots , c_k)\mapsto_{\mathcal{M}} (q', c_1+j_1, \ldots , c_k+j_k)$$
\noi Thus we see that the transition relation must satisfy:
\begin{enumerate}[$-$]
\item
 if $(q, a, i_1, \ldots , i_k, q', j_1, \ldots , j_k) \in \Delta$ and
 $i_m=0$ for some $m\in \{1, \ldots , k\}$, then $j_m=0$ or $j_m=1$
 (but $j_m$ may not be equal to $-1$).
\end{enumerate}

\noindent  
Let $\sigma =a_1a_2 \ldots a_n \ldots $ be an $\om$-word over $\Si$. 
An $\om$-sequence of configurations $r=(q_i, c_1^{i}, \ldots c_k^{i})_{i \geq 1}$ is called 
a run of $\mathcal{M}$ on $\sigma$, starting in configuration 
$(p, c_1, \ldots, c_k)$, iff:
\begin{enumerate}[(1)]
\ite  $(q_1, c_1^{1}, \ldots c_k^{1})=(p, c_1, \ldots, c_k)$

\ite  for each $i\geq 1$, there  exists $b_i \in \Si \cup \{\lambda\}$ such that
 $b_i: (q_i, c_1^{i}, \ldots c_k^{i})\mapsto_{\mathcal{M}}  
(q_{i+1},  c_1^{i+1}, \ldots c_k^{i+1})$  
and such that either ~  $a_1a_2\ldots a_n\ldots =b_1b_2\ldots b_n\ldots$ 
 or ~  $b_1b_2\ldots b_n\ldots$ is a finite prefix of ~ $a_1a_2\ldots a_n\ldots$
\end{enumerate}
\noi The run $r$ is said to be complete when $a_1a_2\ldots a_n\ldots =b_1b_2\ldots b_n\ldots$ 

For every such run, $\mathrm{In}(r)$ is the set of all states entered
 infinitely often during the run $r$.

A complete run $r$ of $M$ on $\sigma$, starting in configuration
 $(q_0, 0, \ldots, 0)$, will be simply called ``a run of $M$ on
 $\sigma$".
\end{defi}

\begin{defi} A B\"uchi $k$-counter automaton  is a 5-tuple 
$\mathcal{M}$=$(K,\Si, \Delta, q_0,$$ F)$, 
where $ \mathcal{M}'$=$(K,\Si, \Delta,$ $ q_0)$
is a $k$-counter machine and $F \subseteq K$ 
is the set of accepting  states.
The \ol~ accepted by $\mathcal{M}$ is 
 $L(\mathcal{M})$= $\{  \sigma\in\Si^\om \mid \mbox{  there exists a  run r
 of } \mathcal{M} \mbox{ on } \sigma \mbox{  such that } \mathrm{In}(r)
 \cap F \neq \emptyset \}$.
\end{defi}

\noi  The class of \ol s accepted by  B\"uchi $k$-counter automata  will be 
denoted ${\bf BCL}(k)_\om$.
 The class of \ol s accepted by {\it  real time} B\"uchi $k$-counter automata  will be 
denoted {\bf r}-${\bf BCL}(k)_\om$.

\hs   Remark that  the $1$-counter automata   introduced above are equivalent to the pushdown automata 
whose stack alphabet is in the form $\{Z_0, A\}$ where $Z_0$ is the bottom symbol which always 
remains at the bottom of the stack and appears only there and $A$ is another stack symbol, see \cite{ABB96}. 

The class ${\bf BCL}(1)_\om$ is  a strict subclass of the class ${\bf CFL}_\om$ of context free \ol s
accepted by B\"uchi pushdown automata.  Notice that an  $\om$-language $L$  is in the class ${\bf BCL}(1)_\om$ 
(respectively, ${\bf CFL}_\om$)  iff it belongs to the $\om$-Kleene closure of the class of 
finitary  languages accepted by $1$-counter automata (respectively, pushdown automata), i.e. 
iff there exist some $1$-counter (respectively, context-free)  languages $U_i, V_i$, for $i\in [1, n] $,  such that 
$L = \bigcup_{i=1}^n U_i.V_i^\om$, see \cite{Staiger97,Fin-mscs06,Fin-survey}. 

\hs We shall consider also the notion of acceptance of binary  relations over infinite words by $2$-tape B\"uchi automata, firstly considered by 
Gire and Nivat in \cite{Gire-Phd,Gire-Nivat}.  A $2$-tape  automaton is an automaton having two tapes and two reading heads, one for each tape, 
which can move asynchronously, and a finite control as in the case of a ($1$-tape) automaton. The automaton reads a pair of (infinite) words 
$(u, v)$ where $u$ is on the first tape and $v$ is on the second tape. Such automata can also be considered for the reading of pairs of finite words 
but we shall only need  here the case of  infinite words. 
We now  recall the formal definition of $2$-tape B\"uchi automata and of  infinitary rational relations. 

\begin{defi}
A  2-tape B\"uchi automaton 
 is a $6$-tuple $\mathcal{T}=(K, \Si_1, \Si_2, \Delta, q_0, F)$, where 
$K$ is a finite set of states, $\Si_1$ and $\Si_2$ are finite  alphabets, 
$\Delta$ is a finite subset of $K \times \Si_1^\star \times \Si_2^\star \times K$ called 
the set of transitions, $q_0$ is the initial state,  and $F \subseteq K$ is the set of 
accepting states. 

 A computation $\mathcal{C}$ of the  
2-tape B\"uchi automaton $\mathcal{T}$ is an infinite sequence of transitions 
$$(q_0, u_1, v_1, q_1), (q_1, u_2, v_2, q_2), \ldots (q_{i-1}, u_{i}, v_{i}, q_{i}), 
(q_i, u_{i+1}, v_{i+1}, q_{i+1}), \ldots $$
\noi The computation is said to be successful iff there exists a final state $q_f \in F$ 
and infinitely many integers $i\geq 0$ such that $q_i=q_f$. 
 The input word of the computation is $u=u_1.u_2.u_3 \ldots$
The output word of the computation is $v=v_1.v_2.v_3 \ldots$
Here the input and the output words may be finite or infinite. 

 The infinitary rational relation $L(\mathcal{T})\subseteq \Si_1^\om
\times \Si_2^\om$ accepted by the 2-tape B\"uchi automaton
$\mathcal{T}$ is the set of pairs $(u, v) \in \Si_1^\om \times
\Si_2^\om$ such that $u$ and $v$ are the input and the output words of
some successful computation $\mathcal{C}$ of $\mathcal{T}$.
\end{defi} 

\begin{Rem}
An infinitary rational relation $L(\mathcal{T})\subseteq \Si_1^\om \times \Si_2^\om$ may be seen as  an $\om$-language over the product alphabet 
$\Si_1 \times \Si_2$. 
\end{Rem}

\hs In the sequel, we will also consider   the notion of recognizable language of infinite pictures.  
We recall first some basic definitions about languages of infinite two-dimensional  words, i.e., languages of infinite pictures. 

\hs Let $\Si$ be a finite alphabet and $\#$ be a letter not in $\Si$ and let 
$\hat{\Si}=\Si \cup \{\#\}$. 
 An $\om$-picture over $\Si$ 
is a function $p$ from $\om \times \om$ into $\hat{\Si}$ such that $p(i, 0)=p(0, i)=\#$ 
for all $i\geq 0$ and $p(i, j) \in \Si$ for $i, j >0$. 
 For each integer $j\geq 1$,  the $j^{th}$ row of the $\om$-picture 
$p$ is the infinite word $p(1, j).p(2, j).p(3, j) \ldots$ over $\Si$ 
and the $j^{th}$ column of $p$ is the infinite word $p(j, 1).p(j, 2).p(j, 3) \ldots$ 
over $\Si$. 
 The set of  $\om$-pictures over $\Si$ is denoted by 
$\Si^{\om, \om}$. An $\om$-picture language $L$ is a subset of $\Si^{\om, \om}$.

\hs In \cite{ATW02}, Altenbernd, Thomas and 
W\"ohrle have considered acceptance of 
languages of infinite two-dimensional words (infinite pictures) by finite tiling systems,  
with the usual acceptance conditions, such as  the 
B\"uchi and Muller ones,  firstly  used for infinite words. They showed that B\"uchi and Muller acceptance conditions lead to the same class 
of recognizable languages of infinite pictures. 
So we shall only recall the notion of B\"uchi recognizable languages of infinite pictures, see \cite{ATW02,Finkel04,Fink-tilings} for more details. 

\hs A tiling system is a tuple $\mathcal{A}$=$(Q, \Si, \Delta)$, where $Q$ is a finite set 
of states, $\Si$ is a finite alphabet, $\Delta \subseteq (\hat{\Si} \times Q)^4$ is a finite set 
of tiles. 

 A B\"uchi tiling system is a pair $(\mathcal{A},$$ F)$ 
 where $\mathcal{A}$=$(Q, \Si, \Delta)$ 
is a tiling system and $F\subseteq Q$ is the set of accepting states. 
 Tiles are denoted by 
\[\left ( \begin{array}{cc}  (a_3, q_3) & (a_4, q_4)
\\ (a_1, q_1) & (a_2, q_2) \end{array} \right )  
\]
  with $a_i\in\hat{\Si}$ and  
  $q_i \in Q$,  and in general, over an alphabet $\Ga$, by 
\[\left ( \begin{array}{cc} b_3 & b_4 
\\ b_1 & b_2 \end{array} \right )  
\]
  with $b_i\in\Ga$.  A combination of tiles is defined  by: 
\[\left ( \begin{array}{cc}  b_3 & b_4
\\ b_1 & b_2 \end{array} \right ) \circ  
\left ( \begin{array}{cc} b'_3 & b'_4
\\ b'_1 & b'_2  \end{array} \right ) = 
\left ( \begin{array}{cc} (b_3, b'_3) & (b_4, b'_4) 
\\ (b_1, b'_1) & (b_2, b'_2) \end{array} \right ) 
\]

\begin{defi} Let $\mathcal{A}$=$(Q, \Si, \Delta)$ 
be a tiling system, and $F\subseteq Q$ be the set of accepting states. 

\noi A run of the  tiling system $\mathcal{A}$=$(Q, \Si, \Delta)$ over an 
$\om$-picture $p \in \Si^{\om, \om}$ 
is a mapping $\rho$  from $\om \times \om$ 
into $Q$ such that for all $(i, j) \in \om \times \om$ 
with $p(i, j)=a_{i, j}$ and $\rho(i, j)=q_{i, j}$ we have 
\begin{displaymath}
\left ( \begin{array}{cc}  a_{i, j+1} & a_{i+1, j+1} 
\\  a_{i, j}  & a_{i+1, j}   \end{array} \right ) \circ  
\left ( \begin{array}{cc} q_{i, j+1} & q_{i+1, j+1} 
\\ q_{i, j}  & q_{i+1, j}   \end{array} \right ) \in \Delta . 
\end{displaymath}

\noi  The $\om$-picture language $L((\mathcal{A},$$ F))$  B\"uchi-recognized 
by $(\mathcal{A},$$ F)$ 
is the set of $\om$-pictures $p \in \Si^{\om, \om}$ such that there is some run $\rho$ of 
$\mathcal{A}$ on $p$ and $\rho(v) \in F$  for infinitely many $v\in \om^2$. 
\end{defi}

\hs An interesting variation of the above defined  reognizability  condition for infinite pictures  uses the  diagonal   of an $\om$-picture. 
The diagonal of an $\om$-picture $p$ is the set of vertices $Di(p)=\{ (i, i) \mid i\in \om \}$. 

\hs The $\om$-picture language B\"uchi-recognized
by $(\mathcal{A},$$ F)$  {\it  on the diagonal} 
is the set of $\om$-pictures $p \in \Si^{\om, \om}$ such that there is some run $\rho$ of 
$\mathcal{A}$ on $p$ and $\rho(v) \in F$ for infinitely many $v\in Di(p)$. 

\hs  The following result was stated  in \cite{ATW02}. 

\begin{thm}
An $\om$-picture language $L \subseteq \Si^{\om, \om}$ is 
 B\"uchi-recognized by a  tiling system if and only if it is 
 B\"uchi-recognized  {\it  on the diagonal} by a  tiling system. \qed
\end{thm}

\noi We can state some  links with classical  notions of tiling of the (quarter of the) plane, see for instance \cite{BJ08-JAC}. 

\hs We denote $\Ga=\hat{\Si} \times Q$ where $\Si$ is the alphabet of pictures and $Q$ is the set of states of a tiling system $\mathcal{A}$=$(Q, \Si, \Delta)$. 
We consider configurations which are elements of $\Ga^{\om \times \om}$. One can imagine that each cell of the quarter of the plane contains a letter 
of the alphabet $\Ga$. 

  Let  $\Delta \subseteq (\hat{\Si} \times Q)^4= \Ga^4$ be  a finite set 
of tiles. We denote its complement   by   $\Delta^-=  \Ga^4 - \Delta$. 
A tiling of  the (quarter of the) plane with $\Delta^-$ as set of  forbidden patterns  is simply a configuration $c\in \Ga^{\om \times \om}$ such that 
for all integers $i, j \in \om$: 
\begin{displaymath}
\left ( \begin{array}{cc}  c(i, j+1) & c(i+1, j+1) 
\\  c(i, j)  & c(i+1, j)   \end{array} \right )   
\in \Delta . 
\end{displaymath}

\noi Then the  $\om$-picture language $L \subseteq \Si^{\om, \om}$ which is  B\"uchi-recognized {\it on the diagonal}
 by the tiling system $(\mathcal{A},$$ F)$
is simply the set of $\om$-pictures $p \in \Si^{\om, \om}$ which are projections of configurations $c\in \Ga^{\om \times \om}$ which are 
tilings of  the (quarter of the) plane with $\Delta^-$ as set of  forbidden patterns  such that 
 for infinitely many $i \in \om$  the second component of  $c(i, i)$ is in $F$.

\section{Topology}

\noi We assume the reader to be familiar with basic notions of topology which
may be found in \cite{Moschovakis80,LescowThomas,Kechris94,Staiger97,PerrinPin}.
There is a natural metric on the set $\Sio$ of  infinite words 
over a finite alphabet 
$\Si$ containing at least two letters which is called the {\it prefix metric} and defined as follows. For $u, v \in \Sio$ and 
$u\neq v$ let $\delta(u, v)=2^{-l_{\mathrm{pref}(u,v)}}$ where $l_{\mathrm{pref}(u,v)}$ 
 is the first integer $n$
such that the $(n+1)^{st}$ letter of $u$ is different from the $(n+1)^{st}$ letter of $v$. 
This metric induces on $\Sio$ the usual  Cantor topology for which {\it open subsets} of 
$\Sio$ are in the form $W.\Si^\om$, where $W\subseteq \Sis$.
A set $L\subseteq \Si^\om$ is a {\it closed set} iff its complement $\Si^\om - L$ 
is an open set.
Define now the {\it Borel Hierarchy} of subsets of $\Si^\om$:

\begin{defi}
For a non-null countable ordinal $\alpha$, the classes ${\bf \Si}^0_\alpha$
 and ${\bf \Pi}^0_\alpha$ of the Borel Hierarchy on the topological space $\Si^\om$ 
are defined as follows:
\begin{enumerate}[$-$]
\item ${\bf \Si}^0_1$ is the class of open subsets of $\Si^\om$, 
\item ${\bf \Pi}^0_1$ is the class of closed subsets of $\Si^\om$, 
\end{enumerate}
 and for any countable ordinal $\alpha \geq 2$: 
\begin{enumerate}[$-$]
\item ${\bf \Si}^0_\alpha$ is the class of countable unions of subsets
of $\Si^\om$ in $\bigcup_{\gamma <\alpha}{\bf \Pi}^0_\gamma$.
\item ${\bf \Pi}^0_\alpha$ is the class of countable intersections of
subsets of $\Si^\om$ in $\bigcup_{\gamma <\alpha}{\bf \Si}^0_\gamma$.
\end{enumerate}
\end{defi}

\noi Recall some basic results about these classes.
 The Borel classes are closed under finite intersections and 
unions, and continuous preimages. Moreover, $\borapxi$ is closed under countable unions, and $\bormpxi$ 
under countable intersections. As usual  the ambiguous class  $\borpxi$ is the class $\borapxi\cap\bormpxi$.

\hs The class of 
{\it Borel\ sets} is $\borel\! :=\!\bigcup_{\xi <\omega_1}\ \borapxi\! =\!
\bigcup_{\xi <\omega_1}\ \bormpxi$, where $\om_1$ is the first uncountable ordinal. The class of Borel sets is the closure of the class of open sets 
under countable union and countable intersection. It is also the closure of the class of open sets under countable union 
(respectively, intersection) and complementation. 

\hs The {\it Borel hierarchy} is as follows:
$$\begin{array}{ll}  
& \ \ \ \ \ \ \ \ \ \ \ \ \ \ \ \ \ \ \ \ \ \ \ \ \ \boraone\! =\!\hbox{\rm open}\ \ \ \ \ \ \ \ \ \ \ \ \ 
\boratwo\! \ \ \ \ \ \ \ \  \ \ \ 
\ldots\ \ \ \ \ \ \ \ \ \ \ \ \boraom\ \ \ \ \ \ldots\cr  
& \borone\! =\!\hbox{\rm clopen}\ \ \ \ \ \ \ \ \ \ \ \ \ \ \ \ \ \ \ \ \ \ \ \ \ \ \ 
\bortwo\ \ \ \ \ \ \ \ \ \ \ \ \ \ \ \ \ \ \ \ \ \ \ \ \ \ \ \ \ \ \ \borom\ \ \ \ \ \ \ \ \ \ \ \ \ \ \ \ \ \ \ \ \ \ \borel\cr
& \ \ \ \ \ \ \ \ \ \ \ \ \ \ \ \ \ \ \ \ \ \ \ \ \ \bormone\! =\!\hbox{\rm closed}\ \ \ \ \ \ \ \ \ \ \bormtwo\! \ \ \ \  \ \ \ \ \ \ \ \ \ldots
\ \ \ \ \ \ \ \ \ \ \ \ \bormom\ \ \ \ \ \ldots
\end{array}$$

\noi This picture means that any class is contained in every class to the right of it, 
and the inclusion is strict in any of the spaces $\Sigma^\omega$.

\noi For 
a countable ordinal $\alpha$,  a subset of $\Si^\om$ is a Borel set of {\it rank} $\alpha$ iff 
it is in ${\bf \Si}^0_{\alpha}\cup {\bf \Pi}^0_{\alpha}$ but not in 
$\bigcup_{\gamma <\alpha}({\bf \Si}^0_\gamma \cup {\bf \Pi}^0_\gamma)$.

\hs    
There are also some subsets of $\Si^\om$ which are not Borel. 
Indeed there exists another hierarchy beyond the Borel hierarchy, which is called the 
projective hierarchy and which is obtained from  the Borel hierarchy by 
successive applications of operations of projection and complementation.
The first level of the projective hierarchy is formed by the class of {\it analytic sets} and the class of {\it co-analytic sets} which are complements of 
analytic sets.  
In particular 
the class of Borel subsets of $\Si^\om$ is strictly included into 
the class  ${\bf \Si}^1_1$ of {\it analytic sets} which are 
obtained by projection of Borel sets. 

\begin{defi} 
A subset $A$ of  $\Si^\om$ is in the class ${\bf \Si}^1_1$ of {\it analytic} sets
iff there exists another finite set $Y$ and a Borel subset $B$  of  $(\Si \times Y)^\om$ 
such that $ x \in A \lra \exists y \in Y^\om $ such that $(x, y) \in B$, 
where $(x, y)$ is the infinite word over the alphabet $\Si \times Y$ such that
$(x, y)(i)=(x(i),y(i))$ for each  integer $i\geq 1$.
\end{defi}

\begin{Rem}
In the above definition we could take $B$ in the class ${\bf \Pi}^0_2$. Moreover 
analytic subsets of $\Si^\om$ are the projections of  ${\bf \Pi}^0_1$-subsets of 
$\Si^\om \times \om^\om$, where $ \om^\om$ is the Baire space, \cite{Moschovakis80}. 
\end{Rem}

\noi By Suslin's Theorem it holds that a subset $A$ of  $\Si^\om$
is Borel iff it is analytic {\it and} coanalytic, i.e. ${\bf \Delta}^1_1 = {\bf \Pi}^1_1 \cap {\bf \Si}^1_1$. 
A set $A$ which is analytic but not coanalytic, or equivalently analytic but not Borel, is called 
a {\it true analytic set}.

\hs   We now define completeness with regard to reduction by continuous functions. 
For a countable ordinal  $\alpha\geq 1$, a set $F\subseteq \Si^\om$ is said to be 
a ${\bf \Si}^0_\alpha$  
(respectively,  ${\bf \Pi}^0_\alpha$, ${\bf \Si}^1_1$)-{\it complete set} 
iff for any set $E\subseteq Y^\om$  (with $Y$ a finite alphabet): 
 $E\in {\bf \Si}^0_\alpha$ (respectively,  $E\in {\bf \Pi}^0_\alpha$,  $E\in {\bf \Si}^1_1$) 
iff there exists a continuous function $f: Y^\om \ra \Si^\om$ such that $E = f^{-1}(F)$. 

 Recall that 
a set $X \subseteq \Sio$ is a ${\bf \Si}^0_\alpha$
 (respectively ${\bf \Pi}^0_\alpha$)-complete subset of $\Sio$ iff it is in ${\bf \Si}^0_\alpha$ 
but not in ${\bf \Pi^0_\alpha}$  (respectively in  ${\bf \Pi}^0_\alpha$ but not in  ${\bf \Si}^0_\alpha$), \cite{Kechris94}. 
 ${\bf \Si}^0_n$
 (respectively ${\bf \Pi}^0_n$)-complete sets, with $n$ an integer $\geq 1$, 
 are thoroughly characterized in \cite{Staiger86a}.

\hs 
In particular, the singletons of $2^\omega$ are $\bormone$-complete subsets of $2^\omega$. 
The $\om$-language   $\mathcal{R}=(0^\star.1)^\om$  
is a well known example of 
${\bf \Pi}^0_2 $-complete subset of $\{0, 1\}^\om$. It is the set of 
$\om$-words over $\{0, 1\}$ having infinitely many occurrences of the letter $1$. 
Its  complement 
$\{0, 1\}^\om - (0^\star.1)^\om$ is a 
${\bf \Si}^0_2 $-complete subset of $\{0, 1\}^\om$.

\hs We recall now the definition of the  arithmetical hierarchy of  \ol s which form the effective analogue to the 
hierarchy of Borel sets of finite ranks. 

 Let $X$ be a finite alphabet. An \ol~ $L\subseteq X^\om$  belongs to the class 
$\Si_n$ if and only if there exists a recursive relation 
$R_L\subseteq (\mathbb{N})^{n-1}\times X^\star$  such that
$$L = \{\sigma \in X^\om \mid \exists a_1\ldots Q_na_n  \quad (a_1,\ldots , a_{n-1}, 
\sigma[a_n+1])\in R_L \}$$
\noi where $Q_i$ is one of the quantifiers $\fa$ or $\exists$ 
(not necessarily in an alternating order). An \ol~ $L\subseteq X^\om$  belongs to the class 
$\Pi_n$ if and only if its complement $X^\om - L$  belongs to the class 
$\Si_n$.  The inclusion relations that hold  between the classes $\Si_n$ and $\Pi_n$ are 
the same as for the corresponding classes of the Borel hierarchy. 
 The classes $\Si_n$ and $\Pi_n$ are  included in the respective classes 
${\bf \Si_n^0}$ and ${\bf \Si_n^0}$ of the Borel hierarchy, and cardinality arguments suffice to show that these inclusions are strict. 

\hs  As in the case of the Borel hierarchy, projections of arithmetical sets 
(of the second $\Pi$-class) lead 
beyond the arithmetical hierarchy, to the analytical hierarchy of \ol s. The first class 
of this hierarchy is the (lightface) class $\Si^1_1$ of {\it effective analytic sets} 
 which are obtained by projection of arithmetical sets.
An \ol~ $L\subseteq X^\om$  belongs to the class 
$\Si_1^1$ if and only if there exists a recursive relation 
$R_L\subseteq \mathbb{N}\times \{0, 1\}^\star \times X^\star$  such that:

$$L = \{\sigma \in X^\om  \mid \exists \tau (\tau\in \{0, 1\}^\om \wedge \fa n \exists m 
 ( (n, \tau[m], \sigma[m]) \in R_L )) \}$$

\noi Then an \ol~ $L\subseteq X^\om$  is in the class $\Si_1^1$ iff it is the projection 
of an \ol~ over the alphabet $X\times \{0, 1\}$ which is in the class $\Pi_2$.  The (lightface)  class $\Pi_1^1$ of  {\it effective co-analytic sets} 
 is simply the class of complements of effective analytic sets. We denote as usual $\Delta_1^1 = \Si^1_1 \cap \Pi_1^1$. 

 Recall that an \ol~ $L\subseteq X^\om$ is in the class $\Si_1^1$
iff it is accepted by a non deterministic Turing machine (reading $\om$-words)
with a   B\"uchi or Muller acceptance condition  \cite{CG78b,Staiger97}.

 \section{The largest thin effective coanalytic set}

\noi We now recall some basic notions of set theory 
which will be useful in the sequel, and which are exposed in any  textbook on set theory, such as  \cite{Jech}.

\hs  The usual axiomatic system {\bf ZFC} is 
Zermelo-Fraenkel system {\bf ZF}   plus the axiom of choice {\bf AC}. 
 A model ({\bf V}, $\in)$ of  the axiomatic system {\bf ZFC} is a collection  {\bf V} of sets,  equipped with 
the membership relation $\in$, where ``$x \in y$" means that the set $x$ is an element of the set $y$, which satisfies the axioms of  {\bf ZFC}.  
We shall often say `` the model {\bf V}"
instead of  ``the model  ({\bf V}, $\in)$". 

\hs The axioms of {\bf ZFC} express some  natural facts that we consider to hold in the universe of sets. For instance a natural fact is that 
two sets $x$ and $y$ are equal iff they have the same elements. 
This is expressed by the sentence: 
$$\fa x \fa y ~ [ ~ x=y \leftrightarrow \fa z ( z\in x \leftrightarrow z\in y ) ~]$$
\noi The above sentence is the {\it Axiom of Extensionality}. 

\hs  Another natural axiom is the {\it Pairing Axiom}   which states that for all sets $x$ and $y$ there exists a  set $z=\{x, y\}$ 
whose elements are $x$ and $y$: 
$$\fa x \fa y ~ [ ~\exists z ( \fa w  ( w\in z \leftrightarrow (w=x \vee w=y) ) ) ]$$

\noi Similarly the {\it Powerset Axiom} states the existence of the set of subsets of a set $x$. 

\hs The Separation Schema is in fact an infinite set of axioms.  For each first-order formula $\varphi$, with free variable $z$, in the language of set theory 
with the equality symbol and the binary symbol $\in$, the following axiom states the existence of the set $y=\{ z\in x \mid \varphi(z) \}$ 
of elements of a set $x$ which satisfy $\varphi$. 
$$\fa x [ \exists y (\fa z ( z\in y \leftrightarrow ( z\in x \wedge \varphi(z) ) ) ) ]$$

\noi The other axioms of {\bf ZFC} are the Union Axiom, the Replacement Schema, the Infinity Axiom, the Foundation Axiom, and the Axiom of Choice. 
We refer the reader to any textbook on set theory, like \cite{Jech},  for an exposition of these axioms.

\hs  We recall that the infinite cardinals are usually denoted by
$\aleph_0, \aleph_1, \aleph_2, \ldots , \aleph_\alpha, \ldots$
The cardinal $\aleph_\alpha$ is also denoted by $\om_\alpha$,
as usual when it is considered as an ordinal.

\hs The continuum hypothesis {\bf CH}  says that the first uncountable cardinal $\aleph_1$ is equal to $2^{\aleph_0}$ which is the cardinal of the 
continuum. G\"odel and Cohen have proved that the continuum hypothesis {\bf CH} is independent from the axiomatic system {\bf ZFC}. This means that there 
are some models of {\bf ZFC + CH} and also some models of {\bf ZFC + $\neg$ CH}, where {\bf $\neg$ CH} denotes the negation of the 
continuum hypothesis, \cite{Jech}. 

\hs 
Let ${\bf ON}$ be the class of all ordinals. Recall that an ordinal $\alpha$ is said to be a successor ordinal iff there exists an ordinal $\beta$ such that 
$\alpha=\beta + 1$; otherwise the ordinal $\alpha$ is said to be a limit ordinal and in that case 
$\alpha ={\rm sup} \{ \beta \in {\bf ON}\mid \beta < \alpha \}$.

\hs  The  class ${\bf L}$ of  {\it constructible sets} in a model {\bf V} of {\bf ZF} is defined by 
$${\bf L} = \bigcup_{\alpha \in {\bf ON}} {\bf L}(\alpha) $$
where the sets ${\bf L}(\alpha) $ are constructed  by induction as follows: 

\begin{enumerate}[(1)]
\ite ${\bf L}(0) =\emptyset$
\ite ${\bf L}(\alpha) = \bigcup_{\beta <  \alpha} {\bf L}(\beta) $, for $\alpha$ a limit ordinal, and 
\ite ${\bf L}(\alpha + 1) $ is the set of subsets of ${\bf L}(\alpha) $ which are definable from a finite number of elements of ${\bf L}(\alpha) $
by a first-order formula relativized to ${\bf L}(\alpha) $. 
\end{enumerate}

\noindent If {\bf V} is a model of {\bf ZF} and ${\bf L}$ is the class
of {\it constructible sets} of {\bf V}, then the class ${\bf L}$ forms
a model of {\bf ZFC + CH}.  Notice that the axiom ({\bf V=L}) means
``every set is constructible" and that it is consistent with {\bf
ZFC}.

\hs Consider now a model {\bf V} of  the axiomatic system {\bf ZFC} and the class of constructible sets ${\bf L} \subseteq {\bf V}$ which forms another 
model of  {\bf ZFC}.  It is known that 
the ordinals of {\bf L} are also the ordinals of  {\bf V}. But the cardinals  in  {\bf V}  may be different from the cardinals in {\bf L}. 

\hs  In the sequel we shall consider in particular the first uncountable cardinal in {\bf L}; it is denoted 
 $\aleph_1^{\bf L}$. It is in fact an ordinal of {\bf V} which is denoted $\om_1^{\bf L}$. 
  It is known that this ordinal satisfies the inequality 
$\om_1^{\bf L} \leq \om_1$.  In a model {\bf V} of  the axiomatic system {\bf ZFC + V=L} the equality $\om_1^{\bf L} = \om_1$ holds. But in 
some other models of {\bf ZFC} the inequality may be strict and then $\om_1^{\bf L} < \om_1$. This is explained in \cite[page 202]{Jech}: one can start 
 from a model 
{\bf V} of {\bf ZFC + V=L} and construct by  forcing  a generic extension {\bf V[G]} in which the cardinals $\om$ and $\om_1$ are 
collapsed; in this extension the inequality $\om_1^{\bf L} < \om_1$ holds.

\hs We now recall the notion of  perfect set. 

\begin{defi} 
Let $P \subseteq \Sio$, where $\Si$ is a finite alphabet having at least two letters. The set 
 $P$ is said to be a perfect subset of $\Sio$ if and only if :  
\begin{enumerate}[(1)]
\item $P$ is a non-empty closed set,  and 
\item for every $x\in P$ and every open set $U$ containing $x$ there
is an element $y \in P\cap U$ such that $x\neq y$.
\end{enumerate}
\end{defi}

\noi So a perfect subset of $\Sio$ is a non-empty closed set which has no isolated points. It is well known that a  perfect subset of 
$\Sio$   has cardinality $2^{\aleph_0}$, i.e. the cardinality of the continuum, see \cite[page 66]{Moschovakis80}. 
We recall now the definition of the {\it perfect set property} and some known results for Borel or analytic sets. 

\begin{defi}
A class ${\bf \Gamma}$ of subsets of $\Sio$ has the perfect set property iff each set $X \in {\bf \Gamma}$ is either countable or contains a perfect subset. 
\end{defi}

\begin{thm}[see \cite{Moschovakis80,Kechris94}]
The class of analytic subsets of $\Sio$ has the perfect set property. In particular, 
 the continuum hypothesis is satisfied for analytic sets: every analytic set is either 
countable or has cardinality  $2^{\aleph_0}$. \qed
\end{thm}

\noi On the other hand,   ``the perfect set property for the class of (effective) coanalytic subsets of $\Sio$" is actually independent from 
the   axiomatic system {\bf ZFC}.  This fact follows easily, as we shall see below, from a result about the 
largest thin effective coanalytic set. 

\hs We first  recall   the notion of  thin  subset of $\Sio$. 

 \begin{defi}
A set $X \subseteq \Sio$ is said to be thin iff it  contains no perfect subset. 
\end{defi}

\noi The important following result was proved by Kechris \cite{Kechris75} and independently by Guaspari \cite{Guaspari}  and  Sacks  \cite{Sacks}. 

\begin{thm}[see \cite{Moschovakis80} page 247]  Let $\Si$ be a finite alphabet having at least two letters. 
There exists a thin $\Pi_1^1$-set $\mathcal{C}_1( \Sio) \subseteq  \Sio$ which contains every thin,  $\Pi_1^1$-subset of $\Sio$. 
It is called the  largest thin $\Pi_1^1$-set  in $\Sio$.  \qed  
\end{thm}

\noi Notice that the existence of the  largest thin $\Pi_1^1$-set  in $\Sio$ is proved from  the   axiomatic system {\bf ZFC}, i.e. 
Zermelo-Fraenkel system {\bf ZF}   plus the axiom of choice {\bf AC}, 
and even if we replace the axiom of choice by a weaker version called the axiom of dependent choice 
{\bf DC}. 

\hs An important fact is that the cardinality of the largest thin $\Pi_1^1$-set in $\Sio$ may depend on the model of {\bf ZFC}.

\hs We can now state Kechris's result on the cardinality of the largest thin $\Pi_1^1$-set, proved independently by Guaspari and Sacks, see also 
\cite[page 171]{Kanamori}. 

\begin{thm}
({\bf ZFC})   The cardinal  of the  largest thin $\Pi_1^1$-set in  $\Sio$ is equal to the cardinal of  $\om_1^{\bf L}$. \qed
\end{thm}

\noi Notice that this means that in a given model {\bf V} of {\bf ZFC} the cardinal  of the  largest thin $\Pi_1^1$-set in  $\Sio$ is equal to the cardinal 
{\it in {\bf V}} of the ordinal $\om_1^{\bf L}$ which plays the role of the cardinal $\aleph_1$ in the inner model {\bf L}  of constructible sets of {\bf V}. 

\hs There exists also a largest thin $\Pi_1^1$-set in  the Baire space $\om^\om$. 
By \cite[Exercise 4F.7, page 251]{Moschovakis80} the cardinal of the  largest thin $\Pi_1^1$-set in  the Baire space is equal to 
the cardinal of the  largest thin $\Pi_1^1$-set in any Cantor space 
 $\Sio$ where $\Si$ is finite and has at least two elements. 

\hs We can now easily state the following result. 

\begin{cor}
The perfect set property for the class of effective coanalytic subsets of $\Sio$ is  independent from 
the   axiomatic system {\bf ZFC}. Indeed it holds that : 
\begin{enumerate}[\em(1)]
\item ({\bf ZFC + V=L}). The class of effective coanalytic subsets of $\Sio$ does not have  the perfect set property. 
\item ({\bf ZFC} + $\om_1^{\bf L} < \om_1$).  The class of effective coanalytic subsets of $\Sio$  has the perfect set property. 
\end{enumerate}
\end{cor}

\proof \hfill
 \begin{enumerate}[(1)]
\item  Assume first that {\bf V} is  a model of the axiomatic system {\bf ZFC + V=L}. In this model 
the cardinal  of the  largest thin $\Pi_1^1$-set in  $\Sio$ is equal to  $\om_1^{\bf L}=\om_1$.  Thus $\mathcal{C}_1( \Sio)$ is not countable but it
contains no perfect subset, hence the  class of effective coanalytic subsets of $\Sio$ does not have  the perfect set property. 

\item
 Assume now that {\bf V} is  a model of the axiomatic system {\bf ZFC} + $\om_1^{\bf L} < \om_1$. In this model the largest 
thin $\Pi_1^1$-set in  $\Sio$ is countable. Thus every  effective coanalytic subset of $\Sio$ is either thin and countable or contains a perfect subset, hence 
the  class of effective coanalytic subsets of $\Sio$  has  the perfect set property. 
\qed 
\end{enumerate}

\noindent Notice that, by \cite[Theorem 14.10, page 184 and   Theorem 11.6, page 136]{Kanamori}, 
 the perfect set property for the class of all (boldface) ${\bf \Pi}_1^1$-subsets of $\Sio$ is 
 equiconsistent with the existence of an 
{\it inaccessible cardinal}, which is a {\it large cardinal}. The axiom ``there exists an inaccessible cardinal" is a ``large cardinal axiom"; its  consistency 
 can not be proved in {\bf ZFC}.  Thus the consistency of the perfect set property for the class of  ${\bf \Pi}_1^1$-subsets of $\Sio$ 
 can not be proved in {\bf ZFC}. We refer the reader to \cite{Kanamori} for an exposition of these results, which will not be necessary 
in this paper. 

\hs On the other hand, if in a model {\bf V} of {\bf ZFC} the class of
${\bf \Pi}_1^1$-subsets of $\Sio$ fails to have the perfect property,
we cannot infer from this property that the continuum hypothesis
is satisfied for ${\bf \Pi}_1^1$-subsets of $\Sio$.  However every
coanalytic set is the union of $\aleph_1$ Borel sets, and this implies
that every coanalytic set is either countable, or of cardinality
$\aleph_1$, or of cardinality $2^{\aleph_0}$, see \cite[Corollary
  25.16, page 488]{Jech}.

\hs We can now state the following results which will be useful in the sequel. 

\begin{cor}\label{cor1}
({\bf ZFC + V=L})    The largest thin $\Pi_1^1$-set in  $\Sio$ is not a Borel set. 
\end{cor}

\proof  In the model {\bf L},  the cardinal of the  largest thin $\Pi_1^1$-set in  $\Sio$ is equal to the cardinal of  $\om_1^{\bf L}$. 
Moreover the continuum hypothesis is satisfied thus $2^{\aleph_0^{{\bf L}}}=\om_1^{\bf L}$. 

  Thus the largest thin $\Pi_1^1$-set in  $\Sio$ has the cardinality 
of the  continuum. But it has no perfect subset and the class of Borel sets has the perfect set property. Thus the  largest thin $\Pi_1^1$-set in  $\Sio$ 
can not be a Borel set. \qed 

\begin{cor}\label{cor2}
({\bf ZFC} + $\om_1^{\bf L} < \om_1$)    The largest thin $\Pi_1^1$-set in  $\Sio$ is countable, hence a ${\bf \Si}^0_2$-set. 
\end{cor}

\proof  Let {\bf V} be a model of  ({\bf ZFC} + $\om_1^{\bf L} < \om_1$).  In this model $\om_1$ is the first uncountable ordinal. Thus 
$\om_1^{\bf L} < \om_1$ implies that $\om_1^{\bf L}$ is a countable ordinal in {\bf V}. Its cardinal is $\aleph_0$ and it is also the cardinal of 
the  largest thin $\Pi_1^1$-set in  $\Sio$.  Thus the set $\mathcal{C}_1( \Sio)$ is countable. But for every $x\in \Sio$ the singleton $\{x\}$ is a closed subset 
of $\Sio$. Thus the largest thin $\Pi_1^1$-set in  $\Sio$ is a countable union of closed sets, i.e. a ${\bf \Si}^0_2$-subset of $\Sio$.
\qed 

\section{Complexity of infinite computations}

\noi  There are several characterizations of the largest thin $\Pi_1^1$-set in $\Sio$, see \cite{Kechris75,Moschovakis80}. 
 Moschovakis gave in   
\cite[page 248]{Moschovakis80} a $\Pi_1^1$-formula $\phi$ defining the  set $\mathcal{C}_1( \Sio)$. Notice that all subformulas of this formula  
are themselves 
given  previously in  the book \cite{Moschovakis80}. 

\hs From now on we shall simply denote by $\mathcal{C}_1$ 
the largest thin $\Pi_1^1$-set in $\{0, 1\}^\om=2^\om$.  

\hs This set  $\mathcal{C}_1$  is a $\Pi_1^1$-set  defined by a $\Pi_1^1$-formula $\phi$. Thus its complement 
$\mathcal{C}_1^-=2^\om - \mathcal{C}_1$ is a   $\Si_1^1$-set  defined by the  $\Si_1^1$-formula $\psi=\neg \phi$. 

\hs  Recall  that one can construct, from the $\Si_1^1$-formula $\psi$ defining $\mathcal{C}_1^-$, 
a B\"uchi Turing machine $\mathcal{T}$ accepting the $\om$-language 
$\mathcal{C}_1^-$, see \cite{Staiger97}. 
 We can then construct  from the  B\"uchi 
Turing machine $\mathcal{T}$, using a classical  construction (see for instance  \cite{HopcroftMotwaniUllman2001}), 
 a $2$-counter  B\"uchi  automaton $\mathcal{A}_1$ accepting the same $\om$-language. 

\hs We are now going to recall some constructions  which were  used in a previous paper \cite{Fin-mscs06} to study topological properties of context-free 
$\om$-languages, and which will be useful in the sequel. 

\hs Let $\Si=\{0, 1\}$,   $E$ be a new letter not in 
$\Si$,  $S$ be an integer $\geq 1$, and $\theta_S: \Sio \ra (\Sigma \cup \{E\})^\om$ be the 
function defined, for all  $x \in \Sio$, by: 
$$ \theta_S(x)=x(1).E^{S}.x(2).E^{S^2}.x(3).E^{S^3}.x(4) \ldots 
x(n).E^{S^n}.x(n+1).E^{S^{n+1}} \ldots $$

\noi We proved in \cite{Fin-mscs06} that if $L \subseteq \Sio$ is an
$\om$-language in the class ${\bf BCL}(2)_\om$ and
$k=cardinal(\Si)+2$, $S=(3k)^3$, then one can construct effectively,
from a B\"uchi $2$-counter automaton $\mathcal{A}_1$ accepting $L$, a
real time B\"uchi $8$-counter automaton $\mathcal{A}_2$ such that
$L(\mathcal{A}_2)=\theta_S(L)$.

\hs We used also in \cite{Fin-mscs06} another coding which we now
recall.  Let $K = 2 \times 3 \times 5 \times 7 \times 11 \times 13
\times 17 \times 19 = 9699690$ be the product of the eight first prime
numbers.  Let $\Ga$ be a finite alphabet; here we shall set
$\Ga=\Si\cup \{E\}$.  An $\om$-word $x\in \Gao$ is coded by the
$\om$-word
\[h_K(x)=A.C^K.x(1).B.C^{K^2}.A.C^{K^2}.x(2).B.C^{K^3}.A.C^{K^3}.x(3).B
 \ldots  B.C^{K^n}. A.C^{K^n}.x(n).B \ldots 
\]
\noi over the alphabet $\Ga_1\!=\!\Ga \cup \{A, B, C\}$, where $A, B, C$
are new letters not in $\Ga$.  In \cite{Fin-mscs06} we proved that,
from a real time B\"uchi $8$-counter automaton $\mathcal{A}_2$
accepting $L(\mathcal{A}_2) \subseteq \Gao$, one can effectively
construct a B\"uchi $1$-counter automaton $\mathcal{A}_3$ accepting
the $\om$-language $h_K( L(\mathcal{A}_2) )$$ \cup h_K(\Ga^{\om})^-$.

\hs Consider now the mapping $\phi_K: (\Ga \cup\{A, B, C\})^\om \ra
 (\Ga \cup\{A, B, C, F\})^\om $ which is simply defined by: for all
 $x\in (\Ga \cup\{A, B, C\})^\om$,
$$\phi_K(x) = F^{K-1}.x(1).F^{K-1}.x(2)  
\ldots F^{K-1}.x(n). F^{K-1}.x(n+1).F^{K-1} \ldots
$$
\noi Then the $\om$-language 
$\phi_K(L(\mathcal{A}_3))=\phi_K ( h_K( L(\mathcal{A}_2) )$$ \cup h_K(\Ga^{\om})^- )$ is accepted by  
 a  real time B\"uchi $1$-counter automaton $\mathcal{A}_4$ which can be effectively 
constructed from the real time B\"uchi $8$-counter automaton $\mathcal{A}_2$, \cite{Fin-mscs06}. 

\hs We can now use these previous constructions to obtain our first main result. 

\hs From now on we consider that we have obtained, from a B\"uchi Turing machine $\mathcal{T}$ accepting the $\om$-language 
$\mathcal{C}_1^-\subseteq \Sio=2^\om$, a $2$-counter  B\"uchi  automaton $\mathcal{A}_1$ accepting the same $\om$-language, and then 
a real time 
B\"uchi $8$-counter automaton $\mathcal{A}_2$ accepting the $\om$-language  $L(\mathcal{A}_2)=\theta_S(\mathcal{C}_1^-)$, 
where $S=(3\times 4)^3=(12)^3$. Next, following the above construction, 
we have a B\"uchi $1$-counter automaton $\mathcal{A}_3$ accepting the $\om$-language  $h_K( L(\mathcal{A}_2) )$$ \cup h_K(\Ga^{\om})^-$, and 
a real time B\"uchi $1$-counter automaton $\mathcal{A}_4$ accepting the $\om$-language  $\phi_K(L(\mathcal{A}_3))$. 
 In the sequel we shall denote simply $\mathcal{A}_4$ by $\mathcal{A}$. 

\begin{thm}\label{mainthe}
 Let $\mathcal{A}$ be the real-time $1$-counter B\"uchi automaton constructed above. The topological complexity of the 
$\om$-language $L(\mathcal{A})$ is not determined by the axiomatic system {\bf ZFC}. Indeed it holds that : 
\begin{enumerate}[\em(1)] 
\item ({\bf ZFC + V=L}). ~~~~~~ The $\om$-language $L(\mathcal{A})$ is a true analytic set. 
\item ({\bf ZFC} + $\om_1^{\bf L} < \om_1$).  ~~~~The $\om$-language $L(\mathcal{A})$ is a  ${\bf \Pi}^0_2$-set. 
\end{enumerate}
\end{thm}

\proof  
\hfill
 \begin{enumerate}[(1)]
\item
 Assume first that {\bf V} is  a model of the axiomatic system {\bf ZFC + V=L}. 
In the model {\bf V},  by Corollary \ref{cor1}  the largest thin $\Pi_1^1$-set   $\mathcal{C}_1$ is  not a Borel set. 
Thus the $\om$-language $\mathcal{C}_1^-=L(\mathcal{A}_1)$ is not a Borel set because the class of Borel subsets of $2^\om$ is closed under 
complementation. The $\om$-language  $L(\mathcal{A}_2)=\theta_S(\mathcal{C}_1^-)$ cannot be a Borel set. Indeed   the function $\theta_S$ is continuous 
and if  $L(\mathcal{A}_2)$ was Borel then the $\om$-language $\mathcal{C}_1^-=\theta_S^{-1}(L(\mathcal{A}_2))$ 
would be Borel too as the inverse image of a Borel set by a continuous function. Next we can see that the $\om$-language 
$L(\mathcal{A}_3)=h_K( L(\mathcal{A}_2) )$$ \cup h_K(\Ga^{\om})^-$ is not Borel. Indeed the function $h_K$ is also continuous and if 
$L(\mathcal{A}_3)$ was Borel then the $\om$-language $L(\mathcal{A}_2)=h_K^{-1}(L(\mathcal{A}_3))$ 
would be Borel too as the inverse image of a Borel set by a continuous function. 
Finally we can see that the $\om$-language $L(\mathcal{A})=\phi_K(L(\mathcal{A}_3))$ is not Borel. Otherwise, the function $\phi_K$ being continuous, 
the $\om$-language $L(\mathcal{A}_3)=\phi_K^{-1}(L(\mathcal{A}))$ would be Borel too as the inverse image of a Borel set by a continuous function. 
Thus the  $\om$-language $L(\mathcal{A})$ is an analytic but non Borel set. 

\item
 Assume now  that {\bf V} is  a model of ({\bf ZFC} + $\om_1^{\bf L} < \om_1$).   In the model {\bf V},  by Corollary \ref{cor2}, 
 the largest thin $\Pi_1^1$-set   $\mathcal{C}_1$ is  a  ${\bf \Si}^0_2$-set. Thus its complement $\mathcal{C}_1^-=L(\mathcal{A}_1)$ is a 
  ${\bf \Pi}^0_2$-set.  It is then proved in \cite{Fin-mscs06} that the $\om$-languages  $L(\mathcal{A}_2)=\theta_S(\mathcal{C}_1^-)$, 
$L(\mathcal{A}_3)=h_K( L(\mathcal{A}_2) )$$ \cup h_K(\Ga^{\om})^-$, and finally $L(\mathcal{A})=\phi_K(L(\mathcal{A}_3))$, are also 
 ${\bf \Pi}^0_2$-sets.\qed 
\end{enumerate}

\noindent We can now improve a recent result from \cite{Fin-HI}. 
 It is very natural to ask whether one can effectively determine the topological complexity of an  $\om$-language accepted by a 
given real-time $1$-counter B\"uchi automaton (respectively, B\"uchi pushdown automaton). 
We had previously shown that this is not possible: For any countable ordinal $\alpha$, it is undecidable whether  an  $\om$-language accepted by a 
given  B\"uchi pushdown automaton is a ${\bf \Si_\alpha^0}$-set (respectively, a ${\bf \Pi_\alpha^0}$-set, a Borel set), \cite{Fin03a}. 
Moreover we have recently  proved  in \cite{Fin-HI} that 
these decision problems are actually $\Pi_2^1$-hard. 
 Notice that here $\Pi_2^1$ is a class of the analytical hierarchy on subsets of  $\mathbb{N}$. The notions
 of analytical hierarchy and of complete sets for classes of this hierarchy may be found for instance in the textbooks  \cite{rog,Odifreddi1,Odifreddi2}.

 A real-time $1$-counter B\"uchi automaton $\mathcal{C}$ has a
 finite description to which can be associated, in an effective way, a
 unique natural number called the index of $\mathcal{C}$. We have then
 a G\"odel numbering of real-time $1$-counter B\"uchi automata.  From
 now on, we shall denote, as in \cite{Fin-HI}, $\mathcal{C}_z$ the
 real time B\"uchi $1$-counter automaton of index $z$ (reading words
 over $\Omega=\{0, 1, A, B, C, E, F\}$).  The above cited result can
 be now formally stated as follows.

\begin{thm}[\cite{Fin-HI}] \label{borel-hard}
\noi Let $\alpha$ be a countable ordinal. Then  
\begin{enumerate}[\em(1)]
\ite $ \{  z \in \mathbb{N}  \mid  L(\mathcal{C}_z) \mbox{ is in the Borel class } {\bf \Si}^0_\alpha \}$ is  $\Pi_2^1$-hard. 
\ite  $ \{  z \in \mathbb{N}  \mid  L(\mathcal{C}_z) \mbox{ is in the Borel class } {\bf \Pi}^0_\alpha \}$ is  $\Pi_2^1$-hard. 
\ite  $ \{  z \in \mathbb{N}  \mid  L(\mathcal{C}_z) \mbox{ is a  Borel set } \}$ is  $\Pi_2^1$-hard. \qed
\end{enumerate} 
\end{thm}

\noi  This implies in particular that these decison problems are not  in the class $\Si^1_2$, but they still  could have been  
$\Pi_2^1$-complete. We are going now to prove that this is not the case. 

\begin{thm} \label{sch}
\noi Let $\alpha$ be a countable ordinal. Then  
\begin{enumerate}[\em(1)]
\ite For $\alpha > 2$,  $ \{  z \in \mathbb{N}  \mid  L(\mathcal{C}_z) \mbox{ is in the Borel class } {\bf \Si}^0_\alpha \}$ is not a  $\Pi_2^1$-set. 
\ite For $\alpha \geq  2$,  $ \{  z \in \mathbb{N}  \mid  L(\mathcal{C}_z) \mbox{ is in the Borel class } {\bf \Pi}^0_\alpha \}$ is not a  $\Pi_2^1$-set. 
\ite  $ \{  z \in \mathbb{N}  \mid  L(\mathcal{C}_z) \mbox{ is a  Borel set } \}$ is  not a  $\Pi_2^1$-set. 
\end{enumerate}
\end{thm}

\proof We first  prove  item (1).  Let   $\mathcal{A}$ be the real-time $1$-counter B\"uchi automaton cited in Theorem \ref{mainthe} and let $z_0$ be its index 
so that $\mathcal{A}=\mathcal{C}_{z_0}$.  

\hs  Assume now  that {\bf V} is  a model of ({\bf ZFC} + $\om_1^{\bf L} < \om_1$).   In the model {\bf V}, by Theorem \ref{mainthe}, the $\om$-language 
$L(\mathcal{A})$  is a  ${\bf \Pi}^0_2$-set, hence also a  ${\bf \Si}^0_\alpha$-set for any countable ordinal $\alpha > 2$. 
Thus, for  $\alpha > 2$, the integer $z_0$ belongs to the set 
$ \{  z \in \mathbb{N}  \mid  L(\mathcal{C}_z) \mbox{ is in the }$$ \mbox{Borel class }$$ {\bf \Si}^0_\alpha \}$. 

\hs But, by Theorem \ref{mainthe}, in the inner model     ${\bf L } \subseteq {\bf V }$, the $\om$-language   $L(\mathcal{A})$  is an analytic but non Borel 
set so  the integer $z_0$ does not belong to the set 
$ \{  z \in \mathbb{N}  \mid  L(\mathcal{C}_z) \mbox{ is in the }$ $ \mbox{Borel class }$ $ {\bf \Si}^0_\alpha \}$. 

\hs On the other hand, Shoenfield's Absoluteness Theorem implies that every  $\Si_2^1$-set (respectively,  $\Pi_2^1$-set)
 is absolute for all inner models of {\rm  (ZF + DC)}, where {\rm  (DC)} is the weak version of the axiom of choice called the 
axiom of dependent choice which
holds in particular in the inner model ${\bf L }$,
see \cite[page 490]{Jech}.

  In particular, if the set   $ \{  z \in \mathbb{N}  \mid  L(\mathcal{C}_z) \mbox{ is in the Borel class } {\bf \Si}^0_\alpha \}$   was a 
$\Pi_2^1$-set, then it could not be a different subset of $\mathbb{N}$ in the models  ${\bf  V}$  and   ${\bf L }$ considered above.
Therefore, for any countable ordinal $\alpha > 2$, 
the   set  $ \{  z \in \mathbb{N}  \mid  L(\mathcal{C}_z) \mbox{ is in the Borel class }$${\bf \Si}^0_\alpha \}$   is not a $\Pi_2^1$-set. 

\hs   Items (2) and (3) follow similarly from Theorem \ref{mainthe} and from Shoenfield's Absoluteness Theorem. 
\qed

\hs In order to prove   similar results for infinitary rational relations accepted by $2$-tape  B\"uchi automata, we shall use a construction from 
\cite{Fin06b}. 
We proved in \cite{Fin06b} that infinitary rational relations have the same topological complexity as $\om$-languages 
accepted by B\"uchi Turing machines. We used a simulation of the behaviour of real time 
$1$-counter automata by $2$-tape B\"uchi automata. 
We recall now a coding which was used in \cite{Fin06b}. 

\hs  We first  define a coding of  an $\om$-word over the finite alphabet $\Omega=\{0, 1,  A, B, C, E, F\}$
by an  $\om$-word over the  alphabet $\Omega' = \Omega \cup \{D\}$, where  $D$ is an additional letter 
not in $\Omega$. 
 For $x\in \Omega^\om$  the $\om$-word $h(x)$ is defined by : 
$$h(x) = D.0.x(1).D.0^2.x(2).D.0^3.x(3).D \ldots D.0^n.x(n).D.0^{n+1}.x(n+1).D \ldots$$
\noi It is easy to see that the mapping $h$ from $\Omega^\om$ into $(\Omega \cup \{D\})^\om$ is continuous and injective. 

\hs Let now  $\alpha$ be the $\om$-word over the alphabet $\Omega'$ 
 which is simply defined by:
$$\alpha = D.0.D.0^2.D.0^3.D.0^4.D \ldots D.0^n.D.0^{n+1}.D \ldots$$

\noi The following results  were  proved in \cite{Fin06b}.

\begin{lem}[\cite{Fin06b}]\label{R1}
Let $\Omega$  be a finite alphabet such that $0\in \Omega$, 
$\alpha$ be the $\om$-word over  $\Omega \cup \{D\}$ defined as above, and 
 $L \subseteq \Omega^\om$ be in  {\bf r}-${\bf BCL}(1)_\om$.
Then there exists  an infinitary rational relation 
$R_1 \subseteq  (\Omega \cup \{D\})^\om \times (\Omega \cup \{D\})^\om$ such that:
$$\fa x\in \Omega^{\om}~~~ (x\in L) \mbox{  iff } ( (h(x), \alpha) \in R_1 )$$ \qed
\end{lem}

\begin{lem}[\cite{Fin06b}]\label{complement} The set 
$R_2 = (\Omega \cup \{D\})^\om \times (\Omega \cup \{D\})^\om - ( h(\Omega^{\om}) \times \{\alpha\} )$
 is an infinitary rational relation.  \qed
\end{lem}

\noi Considering the union $R_1 \cup R_2$ of the two infinitary rational relations obtained in the two above lemmas we get the following result. 

\begin{Pro}[\cite{Fin06b}]\label{pro-ratrel}  Let  $L \subseteq \Omega^\om$ be in  {\bf r}-${\bf BCL}(1)_\om$ and $\mathcal{L}= h(L)  \cup (h(\Omega^{\om}))^- $.  
 Then 
 $$R = \mathcal{L} \times \{\alpha\} ~~ \bigcup  ~~(\Omega')^\om \times ( (\Omega')^\om - \{\alpha\})$$
\noi is an  infinitary rational relation. 
 Moreover  one can construct effectively, from a real time $1$-counter B\"uchi automaton $\mathcal{A}$ accepting $L$, 
a $2$-tape B\"uchi automaton $\mathcal{B}$ accepting the infinitary relation $R$. \qed
\end{Pro}

\noi Let now  $\mathcal{A}$ be the real time $1$-counter B\"uchi automaton constructed  above and cited in Theorem \ref{mainthe} and 
$\mathcal{B}$ be the $2$-tape B\"uchi automaton which can be constructed from  $\mathcal{A}$ by the above Proposition \ref{pro-ratrel}.
 We can now state our second main result. 

\begin{thm}\label{mainthe2}
 The topological complexity of the 
 infinitary rational relation  $L(\mathcal{B})$ is not determined by the axiomatic system {\bf ZFC}. Indeed it holds that : 
\begin{enumerate}[\em(1)]
\item ({\bf ZFC + V=L}). ~~~~~~ The relation $L(\mathcal{B})$ is a true analytic  set. 
\item ({\bf ZFC} + $\om_1^{\bf L} < \om_1$).  ~~~~The relation $L(\mathcal{B})$ is a  ${\bf \Pi}^0_2$-set. 
\end{enumerate}
  \end{thm}

\proof 
\hfill
 \begin{enumerate}[(1)]
\item
Assume first that {\bf V} is  a model of the axiomatic system {\bf ZFC + V=L}. 
In the model {\bf V},  by Corollary \ref{cor1}  the largest thin $\Pi_1^1$-set   $\mathcal{C}_1$ is  not a Borel set and by  Theorem \ref{mainthe}
 the  $\om$-language $L(\mathcal{A})$ is a true analytic set.  

 On the other hand the function $h$ is continuous. Thus the function $g$ from $\Omega^\om$ into 
$(\Omega \cup \{D\})^\om \times (\Omega \cup \{D\})^\om$ defined by $g(x)=(h(x), \alpha)$ is also continuous. If the relation $L(\mathcal{B})$ 
was a Borel set then the $\om$-language $L(\mathcal{A})=g^{-1}(L(\mathcal{B}))$ would be also a Borel set as the inverse image of a Borel set by a 
continuous function. 
Thus the relation $L(\mathcal{B})$ is not a  Borel set. 

\item
 Assume now  that {\bf V} is  a model of ({\bf ZFC} + $\om_1^{\bf L} < \om_1$).   In the model {\bf V},  by Corollary \ref{cor2}, 
 the largest thin $\Pi_1^1$-set   $\mathcal{C}_1$ is  a  ${\bf \Si}^0_2$-set  and  by  Theorem \ref{mainthe} the $\om$-language 
$L(\mathcal{A})$ is a  ${\bf \Pi}^0_2$-set. It is easy to prove that $\mathcal{L}= h(L(\mathcal{A}))  \cup (h(\Omega^{\om}))^- $  is also a 
${\bf \Pi}^0_2$-set (this is due to the fact that $h$ is an homeomorphism between $\Omega^\om$ and its image $h(\Omega^{\om})$ 
which is a closed subset of  $(\Omega \cup \{D\})^\om$, see  \cite{Fin06b}).  Then one can easily see that the set  $\mathcal{L} \times \{\alpha\}$ is also 
a ${\bf \Pi}^0_2$-set. But the set  $(\Omega')^\om \times ( (\Omega')^\om - \{\alpha\})$  is an open hence ${\bf \Pi}^0_2$-subset of 
$(\Omega \cup \{D\})^\om \times (\Omega \cup \{D\})^\om$. 
Thus the relation $R = \mathcal{L} \times \{\alpha\} ~~ \bigcup  ~~(\Omega')^\om \times ( (\Omega')^\om - \{\alpha\})$ is a 
${\bf \Pi}^0_2$-subset of $(\Omega \cup \{D\})^\om \times (\Omega \cup \{D\})^\om$.
\qed 
\end{enumerate}

\noindent From now on we shall denote $\mathcal{T}_z$ the 
$2$-tape B\"uchi automaton of index $z$. Then we recall the following  recent result  which shows that 
 topological properties of infinitary rational relations are  highly undecidable. 

\begin{thm}[ \cite{Fin-HI}]
\noi Let $\alpha$ be a non null countable ordinal. Then  
\begin{enumerate}[\em(1)]
\ite $ \{  z \in \mathbb{N}  \mid  L(\mathcal{T}_z) \mbox{ is in the Borel class } {\bf \Si}^0_\alpha \}$ is  $\Pi_2^1$-hard. 
\ite  $ \{  z \in \mathbb{N}  \mid  L(\mathcal{T}_z) \mbox{ is in the Borel class } {\bf \Pi}^0_\alpha \}$ is  $\Pi_2^1$-hard. 
\ite  $ \{  z \in \mathbb{N}  \mid  L(\mathcal{T}_z) \mbox{ is a  Borel set } \}$ is  $\Pi_2^1$-hard. \qed
\end{enumerate}
\end{thm}

\noindent We can now state that these decision problems are not in the class $\Pi_2^1$. 

\begin{thm} \label{borel-hard-2}
\noi Let $\alpha$ be a countable ordinal. Then  
\begin{enumerate}[\em(1)]
\ite For $\alpha > 2$,  $ \{  z \in \mathbb{N}  \mid  L(\mathcal{T}_z) \mbox{ is in the Borel class } {\bf \Si}^0_\alpha \}$ is not a  $\Pi_2^1$-set. 
\ite For $\alpha \geq  2$,  $ \{  z \in \mathbb{N}  \mid  L(\mathcal{T}_z) \mbox{ is in the Borel class } {\bf \Pi}^0_\alpha \}$ is not a  $\Pi_2^1$-set. 
\ite  $ \{  z \in \mathbb{N}  \mid  L(\mathcal{T}_z) \mbox{ is a  Borel set } \}$ is  not a  $\Pi_2^1$-set. 
\end{enumerate}
\end{thm}

\proof We can reason as  in 
the proof of Theorem \ref{sch} (in the case of  $\om$-languages of $1$-counter B\"uchi automata).  
We use Shoenfield's Absoluteness Theorem and Theorem \ref{mainthe2} instead of  Theorem \ref{mainthe}. 
\qed

\hs We consider now  B\"uchi recognizable languages of infinite pictures.
We shall  use in the sequel a result proved in  \cite{Finkel04,Fink-tilings} which we now recall. 

\hs For $\sigma \in \Sio=\{0, 1\}^\om$ we denote $\sigma^0$ the $\om$-picture whose first row is the $\om$-word $\sigma$ and whose other rows 
are labelled with the letter $0$. 
 For an \ol~  $L \subseteq \Sio=\{0, 1\}^\om$  we  denote  $L^0$  the language of infinite pictures $ \{ \sigma^0 \mid \sigma \in L \}$.

\begin{lem}[\cite{Finkel04}] \label{lemTM}
If   $L \subseteq \Sio$ is 
accepted by some Turing machine with a B\"uchi acceptance 
condition, then $L^0$ is B\"uchi recognizable by a finite tiling system. \qed
\end{lem}

\noi  Recall that for $\Ga$ a finite alphabet having at least two letters, the 
set $\Ga^{\om \times \om}$ of functions  from $\om \times \om$ into $\Ga$ 
is usually equipped with the product topology  of the discrete 
topology on $\Ga$.
This topology may be defined 
by the following distance $d$. Let $x$ and $y$  in $\Ga^{\om  \times \om}$ 
such that $x\neq y$, then  
$$ d(x, y)=\frac{1}{2^n}\ ,\quad\mbox{where}\quad
n=min\{p\geq 0 \mid  \exists (i, j) ~~ x(i, j)\neq y(i, j) \mbox{ and } i+j=p\}.$$
\noi Then the topological space $\Ga^{\om \times \om}$ is homeomorphic to the 
topological space $\Ga^{\om}$, equipped with the Cantor topology.  

  The set $\Si^{\om, \om}$ of $\om$-pictures over $\Si$, 
viewed as a topological subspace of $\hat{\Si}^{\om \times \om}$, 
 is easily seen to be homeomorphic to the topological space $\Si^{\om \times \om}$, 
via the mapping 
$\varphi: \Si^{\om, \om} \ra \Si^{\om \times \om}$ 
defined by $\varphi(p)(i, j)=p(i+1, j+1)$ for all 
$p\in \Si^{\om, \om}$ and $i, j \in \om$. 

\hs  Let now $\mathcal{T}$ be a B\"uchi Turing machine accepting the $\om$-language $\mathcal{C}_1^-$. Using Lemma \ref{lemTM} we can construct a 
B\"uchi tiling system  $\mathcal{S}$ accepting the $\om$-picture language $(\mathcal{C}_1^-)^0$. 
 We consider now the topological complexity of this set $L(\mathcal{S}) \subseteq \Si^{\om, \om}$

\hs It is then easy to see that if $L \subseteq \Sio=\{0, 1\}^\om$ is a ${\bf \Pi}^0_2$-subset of $\Sio$ then the $\om$-picture language 
 $L^0$ is a ${\bf \Pi}^0_2$-subset of $\Si^{\om, \om}$. And if $L \subseteq \Sio=\{0, 1\}^\om$ is not Borel then the $\om$-picture language 
$L^0$ is also not Borel. 
 Then  Corollaries  \ref{cor1} and  \ref{cor2}  imply the following result. 

\begin{thm}\label{mainthe3}
 The topological complexity of the $\om$-picture language $L(\mathcal{S})$
   is not determined by the axiomatic system {\bf ZFC}. Indeed it holds that : 
\begin{enumerate}[\em(1)]
\item ({\bf ZFC + V=L}). ~~~~~~ The $\om$-picture language $L(\mathcal{S})$ is a true analytic set. 
\item ({\bf ZFC} + $\om_1^{\bf L} < \om_1$).  ~~~~The $\om$-picture language $L(\mathcal{S})$ is a  ${\bf \Pi}^0_2$-set. \qed
\end{enumerate}
\end{thm}

\noindent We have recently proved that the topological complexity of  $\om$-picture languages accepted by B\"uchi tiling systems is highly undecidable. 
Below  the  B\"uchi tiling system of index $z$ is denoted by $\mathcal{S}_z$. 

\begin{thm}[ \cite{Fink-tilings}]
\noi Let $\alpha$ be a non null countable ordinal. Then  
\begin{enumerate}[\em(1)]
\ite $ \{  z \in \mathbb{N}  \mid  L(\mathcal{S}_z) \mbox{ is in the Borel class } {\bf \Si}^0_\alpha \}$ is  $\Pi_2^1$-hard. 
\ite  $ \{  z \in \mathbb{N}  \mid  L(\mathcal{S}_z) \mbox{ is in the Borel class } {\bf \Pi}^0_\alpha \}$ is  $\Pi_2^1$-hard. 
\ite  $ \{  z \in \mathbb{N}  \mid  L(\mathcal{S}_z) \mbox{ is a  Borel set } \}$ is  $\Pi_2^1$-hard. \qed
\end{enumerate}
\end{thm}

\noindent As in the case of $\om$-languages of $1$-counter automata or of $2$-tape automata, we can now infer the following result from 
 Shoenfield's Absoluteness Theorem and Theorem \ref{mainthe3}.

\begin{thm} \label{borel-hard-3}
  Let $\alpha$ be a countable ordinal. Then  
\begin{enumerate}[(1)]
\ite For $\alpha > 2$,  $ \{  z \in \mathbb{N}  \mid  L(\mathcal{S}_z) \mbox{ is in the Borel class } {\bf \Si}^0_\alpha \}$ is not a  $\Pi_2^1$-set. 
\ite For $\alpha \geq  2$,  $ \{  z \in \mathbb{N}  \mid  L(\mathcal{S}_z) \mbox{ is in the Borel class } {\bf \Pi}^0_\alpha \}$ is not a  $\Pi_2^1$-set. 
\ite  $ \{  z \in \mathbb{N}  \mid  L(\mathcal{S}_z) \mbox{ is a  Borel set } \}$ is  not a  $\Pi_2^1$-set. \qed
\end{enumerate}
\end{thm}

\section{Concluding remarks}

\noi We obtained  surprising results which show  that 
 the topological complexity of an $\om$-language accepted by a $1$-counter B\"uchi automaton, of an infinitary rational relation 
accepted by a $2$-tape B\"uchi automaton, or of a  B\"uchi recognizable language of infinite pictures, 
is not determined by the  axiomatic system {\bf ZFC}. 

\hs We have inferred from the proof of the above results and from  Shoenfield's Absoluteness Theorem
an improvement of  the lower bound of some decision problems recently studied in \cite{Fin-HI,Fink-tilings}. 

\hs Recall that, by \cite[Remark 3.25]{Fin-HI}, if $\alpha$ is an
ordinal smaller than the Church-Kleene ordinal $\om_1^{\mathrm{CK}}$,
which is the first non-recursive ordinal, then $\{ z \in \mathbb{N}
\mid L(\mathcal{C}_z) \mbox{ is in the Borel }$ $\mbox{class } {\bf
  \Si}^0_\alpha \}$ (respectively, $\{ z \in \mathbb{N} \mid
L(\mathcal{C}_z) \mbox{ is in}$ $\mbox{the Borel class }$ $ {\bf
  \Pi}^0_\alpha \}$) is a $\Si_3^1$-set.  We now know that for $\alpha
> 2$ (respectively, $\alpha \geq 2$), it is actually in the class
$\Si_3^1 \setminus (\Si_2^1 \cup \Pi_2^1)$ but the question is still
open whether these problems are $\Si_3^1$-complete.  The exact
complexity of being in the Borel class ${\bf \Si}^0_\alpha $
(respectively, ${\bf \Pi}^0_\alpha $), for a countable ordinal
$\alpha$, remains an open problem for $\om$-languages of real time
$1$-counter automata (respectively, pushdown automata, $2$-tape
automata) and for B\"uchi recognizable languages of infinite pictures.

\section*{Acknowledgement}

\noi I wish to thank   the  referees
for their very useful comments on a preliminary version of this paper.

\end{document}